\documentclass[prb, preprint]{revtex4}
\usepackage{graphicx}
\usepackage{epsfig}

\begin{document}

\title{Optimized parallel tempering simulations of proteins}

\author{Simon Trebst$^{1,2,3}$, Matthias Troyer$^{2}$,  Ulrich H.E. Hansmann$^{4,5}$}

\affiliation{$^{(1)}$Computational Laboratory,  Eidgen\"ossische Technische Hochschule
 Z\"urich, CH-8092 Z\"urich, Switzerland}
\affiliation{$^{(2)}$Theoretische Physik, Eidgen\"ossische Technische Hochschule
 Z\"urich, CH-8093 Z\"urich, Switzerland}
\affiliation{$^{(3)}$Microsoft Research and Kavli Institute for Theoretical Physics, University of California, Santa Barbara, CA 93106, USA}
\affiliation{$^{(4)}$Department of Physics, Michigan Technological University, Houghton, MI 49931, USA}
\affiliation{$^{(5)}$John-von-Neumann Institute for Computing, Forschungszentrum J\"ulich, D-52425 J\"ulich, Germany}
\date{\today}

\begin{abstract}
We apply a recently developed adaptive algorithm that systematically improves the efficiency of 
parallel tempering or replica exchange methods in the numerical simulation of small proteins. 
Feedback iterations allow us to identify an optimal set of temperatures/replicas which are
found to concentrate at the bottlenecks of the simulations. 
A measure of convergence for the equilibration of the parallel tempering algorithm is discussed. 
We test our algorithm by simulating the 36-residue villin headpiece sub-domain HP-36 where 
we find a lowest-energy configuration with a root-mean-square-deviation of less than $4$ \AA\ 
to the experimentally determined structure. \\ \newline

%\noindent
%{\bf Corresponding author}\\
%Ulrich H.E. Hansmann \\
%John-von-Neumann Institute for Computing \\
%Forschungszentrum J\"ulich \\ 
%D-52425 J\"ulich, Germany \\
%Phone:	+49-2461-61-1526 \\
%Fax:		+49-2461-61-2430 \\ 
%E-mail:	u.hansmann@fz-juelich.de \\ \newline \newlinex
%{\bf Manuscript information}\\
%text pages: 18\\
%words in abstract: 90\\
%characters in paper: approx. 40'000\\
\end{abstract}

\maketitle

% Introduction ---------------------------------------------------------------------
\section{Introduction}
Understanding the folding of proteins from computer simulations is
a longstanding but still elusive goal in computational biology. The 
difficulties stem from the fact that proteins are only marginal stable. 
At room temperature the free energy difference between the 
biologically active and unfolded states is only of order $\approx 10$ 
Kcal/mol. However, this small gap is due to cancelations of large 
energetic and entropic terms which poses two major challenges to
numerical simulations. On the one hand, one has to find a universal 
model that captures this delicate balance. On the other hand, the 
competing interactions necessarily lead to a rugged energy 
landscape that makes the exhaustive sampling of low-temperature 
configurations a challenging computational task. In general, it has 
been hard to distinguish which of the two difficulties is the limiting
factor in computational protein studies.

In this paper we address the second challenge and apply a 
powerful sampling technique that allows to efficiently explore a 
complex energy landscape by systematically shifting computational
resources towards the bottlenecks of a simulation \cite{OptimalEnsemble,OptimalTempering}, 
which are typically in the vicinity of free energy barriers.
We test our algorithm by simulating the 36-residue villin headpiece 
sub-domain HP-36. This molecule has raised considerable interest in 
computational biology \cite{Freed,Ripoli}  as it is one of the smallest 
proteins with well-defined secondary and tertiary structure \cite{McKnight:96}  
but at the same time with 596 atoms still accessible to simulations 
\cite{Duan:98}. Its structure which was resolved by NMR analysis and 
deposited in the Protein Data Bank (PDB code 1vii) is shown in Fig.~1.

Recent computationally intensive investigations have studied this protein
using molecular dynamics \cite{Zagrovic:02} and parallel tempering \cite{Li:03} 
techniques. While the former study reports room temperature configurations
that are within $<4.0$ \AA\ to the native structure, these randomly sampled 
configurations could not be singled out from the misfolded structures in a 
rigorous way. The latter study \cite{Li:03} tries to identify the 
biologically active state as those configurations which minimize the energy 
functional of an implicit solvent model. However, despite considerable long 
simulation times low-energy configurations that resemble the experimentally 
determined one were found  only with less than 20\% frequency at T=250 K. 
These conformers differed still by root-mean-square-deviations (RMSD) 
of $\sim 4- 6$\AA\  from the native structure and could also not be distinguished
 by their energies from that of the predominant misfolded structure.
% NMR configuration
%\begin{figure}[b]
%  \includegraphics[width=0.5\columnwidth]{hp36.eps}
%  \caption{(color online) NMR-derived structure of the 36-residue peptide HP-36 
%                   as deposited in the Protein Data Bank (1vii).}
%  \label{Fig:HP36}
%\end{figure}
What leads to the discrepancy with the experiments? 
The authors of Ref.~\onlinecite{Li:03} argue that it is due to poor approximations of 
the simulated force field and especially the implicit solvent model. Indeed, 
configurations with an RMSD of   $\approx 4$ \AA\  have been found later 
with high frequency in simulations with a  modified energy function \cite{Hansmann:04}. 
However, the alterations of the  implicit solvent model are ad hoc and  not 
universal \cite{WH:05}, while the parameters of the original model were fitted 
against experimental data. 
On the other hand, the data of Ref.~\onlinecite{Li:03} could also indicate that
despite large computational efforts the simulation has not thermalized
and the correct equilibrium distribution of low-energy configurations
not yet been found.

Deciding between the two alternatives in the above example is especially 
important as parallel tempering ~\cite{Tempering} (also known as replica 
exchange method) has recently become the simulation technique of choice 
in protein studies~\cite{Hansmann:97,Hansmann:99,Hansmann:03}.
The question can be re-formulated as how does one gauge the efficiency
of a parallel tempering run and ensures that the sampling is sufficiently long
to ensure thermal equilibration?
The present paper describes  a measure for this purpose and discusses a 
protocol that allows one to optimize the performance of parallel tempering runs 
by finding the best temperature distribution. Using the enhanced parallel
tempering protocol we demonstrate that the simulation of  Ref.~\onlinecite{Li:03}
had indeed not thermalized. On the contrary, we now find a {\em dominant} lowest 
energy configuration that is within $<4.0$ \AA\ to the native structure. 
This RMSD is comparable to the best ones found in previous molecular dynamics
simulations \cite{Zagrovic:02} with a different search technique and energy
function, but our approach requires only 1\% of their computational resources.

% Simulations ---------------------------------------------------------------------

\section{Algorithm Design}
% Parallel tempering simulations ---------------------------------------------------------------------
In parallel tempering \cite{Tempering} simulations $N$ non-interacting copies, 
or ``replicas'', of the protein are simultaneously simulated at a range of temperatures 
$\{T_1, T_2, \ldots, T_N\}$, e.g. by distributing the simulation over $N$ nodes of a 
parallel computer.
After a fixed number of Monte Carlo sweeps (or a molecular dynamics run of a certain 
time) a sequence of swap moves, the exchange of two replicas at neighboring 
temperatures, $T_i$ and $T_{i+1}$, is suggested and accepted with a probability
\begin{equation}
  p(E_i, T_i \rightarrow E_{i+1}, T_{i+1}) = \min \left( 1,  \exp( -\Delta\beta \Delta E) \right)\;,
\end{equation}
where $\Delta\beta = 1/T_{i+1} - 1/T_i$ is the difference between the inverse temperatures 
and $\Delta E = E_{i+1} -E_i$ is the difference in energy of the two replicas. For a given 
replica the swap moves induce a random walk in temperature space that allows the replica 
to wander from low temperatures, where barriers in a complex  energy landscape lead to
long relaxation times, to high temperatures, where equilibration is rapid, {\it and back}.
%equilibration is rapid, but also to  return to low temperatures 
%where relaxation times are suppressed by barriers in a complex energy landscape. 
The convergence of a parallel tempering run is given by the relaxation at lowest 
temperature and can be gauged by the frequency of statistically independent visits at this 
temperature. A lower bound for this number is the rate of round-trips $n_{rt}$ between 
the lowest and highest temperature, $T_1$ and $T_N$ respectively. An equivalent measure 
is the round-trip time $\tau_{1N}$, i.e. the average time it takes a replica to move from lowest 
temperature $T_1$ to the highest  temperature $T_N$, and back. It is this {\em non-local} 
measure in temperature space that one has to minimize in order to optimize a parallel 
tempering simulation \cite{Dayal:04,OptimalEnsemble}. 
Commonly, it is assumed that equilibration is fastest if the {\em local} acceptance rate of 
swaps is the same for all pairs of neighboring temperatures $T_i$ and $T_{i+1}$
\cite{Kofke:04,Predescu:04,Kone:05,Rathore:05,Predescu:05}. 
Recently, it has been shown that this assumption is misleading \cite{OptimalTempering}.
Here we review the algorithm outlined in Ref.~\cite{OptimalTempering} and apply it to
systematically optimize the temperature set used in our simulations in such a way that for 
each replica the number of round-trips is maximized, and equilibration of the system at 
low temperatures thereby substantially improved. 

We illustrate this approach by an example parallel tempering simulation of the 36-residue 
protein HP-36 in an all-atom representation. The intramolecular interactions are described 
by the ECEPP/2 force field \cite{ECEPP} 
\begin{eqnarray}
  E_{\text{ECEPP/2}} &=& E_C + E_{LJ}  + E_{tor} \nonumber \\
  &=&  \sum_{(i,j)} \frac{332 q_i q_j}{\epsilon r_{ij}} + \sum_{(i,j)} \left( \frac{A_{ij}}{r_{ij}^{12}} - \frac{B_{ij}}{r_{ij}^6} \right) \nonumber \\ 
  && + \sum_l U_l ( 1\pm \cos(n_l \xi_l)) \;,
\end{eqnarray}
where $r_{ij}$ is the distance between the atoms $i$ and $j$,  $\xi_l$ is the $l$-th torsion 
angle, and energies are measured in Kcal/mol.  
The protein-solvent interactions are approximated by a solvent accessible surface term 
\begin{equation}
  E_{solv} = \sum_i \sigma_i A_i \;.
\end{equation}
Here $A_i$ is the  solvent accessible surface area of the $i-th$ atom in a given
configuration, and $\sigma_i$ a solvation parameter for the atom $i$. For the 
present investigation we use the  parameter set OONS of Ref.~\onlinecite{OONS}.
Our implementation is based on the software package SMMP 
({\em S}imple {\em M}olecular {\em M}echanics for {\em P}roteins) \cite{SMMP} 
which allowed us to distribute the simultaneous simulation of $N=20$ replicas on 
a beowulf cluster with 2.2~GHz Opteron processors.
%Note that the implementation of the force field differs slightly from the one of 
%Ref.~\cite{Li:03} leading to differences in the absolute energy values.
The initial temperature distribution for these replicas is listed in Table~\ref{Table:TSets}. 
A sequence of swap moves between neighboring temperatures is attemped after each 
Monte Carlo sweep where a sweep consists of a series of Metropolis tests for each of the 
dihedral angles. Note that the implementation of  the force field differs slightly from the one 
in Ref.~\onlinecite{Li:03} leading to (irrelevant) differences in the absolute energy values.

Our approach to optimize the simulated temperature set is inspired by a recently 
introduced adaptive broad-histogram algorithm \cite{OptimalEnsemble} that maximizes 
the rate of round-trips in energy space by shifting additional weight toward the bottlenecks   
of  the simulation and has been outlined in the context of classical spin models in Ref.~\cite{OptimalTempering}. The bottlenecks of the simulation are identified by measuring the local 
diffusivity of the simulated random walk. 
In the case of a parallel tempering run, where we simulate a random 
walk in temperature space, this quantity is calculated by adding a label ``up" or ``down"  to the 
replica that indicates which of the two extremal temperatures, $T_1$ or $T_N$ respectively, 
the replica has visited most recently. The label of a replica changes 
only when the replica visits the opposite extremum. For instance, the label of a replica with 
label ``up"  remains unchanged if the replica comes back to the lowest temperature $T_1$, 
but changes to ``down" upon its first visit to $T_N$. 
For each temperature point in the temperature set $\{ T_i\}$ we record two histograms 
$n_{\rm up} (T_i)$ and $n_{\rm down}(T_i)$. Before attempting a sequence of swap moves we 
increment  at temperature $T_i$ that of the two histograms which has the label of the respective replica 
currently at temperature $T_i$. If a replica has not yet visited neither of the two extremal 
temperatures, we increment neither of the histograms.
For each temperature point this allows us to evaluate the average fraction of replicas which diffuse
from the lowest to the highest temperature as
\begin{equation}
f(T) = \frac{n_{\rm up}(T)}{n_{\rm up}(T)+n_{\rm down}(T)}.
\end{equation} 
In Fig.~\ref{Fig:Fraction} this fraction is plotted for our parallel tempering simulations of HP-36
with an initial temperature distribution as listed in Table \ref{Table:TSets}. 

% fraction & feedback
%\begin{figure}[t]
%  \includegraphics[width=0.75\columnwidth]{Feedback.eps}
%  \caption{Fraction of replicas diffusing from the lowest, $T_1=250~$K, to the highest temperature
%                  $T_N=1000~$K in a parallel tempering simulation of HP-36.
%                  For the optimized temperature set (iteration 3), the temperature points are distributed in
%                  such a way that the fraction shows a nearly constant decay 
%                  $\Delta f_i = f(T_i) - f(T_{i+1})=1/(N-1)$ between adjacent temperature points,
%                  i.e. $\Delta f_i (N-1) \approx const$.}
%  \label{Fig:Fraction}
%\end{figure}

The so-labelled replicas define a steady-state current current from $T_1$ to $T_N$ that is proportional to the round-trip rate $n_{rt}$ and therefore independent of temperature. To first order in the derivative this current is given by
\begin{equation}
   j = D(T) \eta(T) \frac{df}{dT} \;,
   \label{Eq:Current}
\end{equation}  
where $D(T)$ is the local diffusivity at temperature $T$ and $\eta (T)$ is the probability distribution for a replica to reside at temperature $T$, where the temperature $T$ is now assumed to be a continuous variable (and not limited to the points of the current temperature set).
For a given temperature set we approximate this probability distribution with a
step-function $\eta(T) = C /\Delta T$, where $\Delta T = T_{i+1} - T_i$ is the length of the temperature interval around temperature $T_i < T < T_{i+1}$ in the current temperature set. The normalization constant $C$ is chosen as
\begin{equation}
  \int_{T_1}^{T_N} \eta(T) dT = C \int_{T_1}^{T_N} \frac{dT}{\Delta T} = 1 \;.
  \label{Eq:Normalization}
\end{equation}
Rearranging Eq.~(\ref{Eq:Current}) gives a simple measure of the local diffusivity $D(T)$:
\begin{equation}
  D(T) \sim \frac{\Delta T}{df/dT} \;,
  \label{Eq:Diffusivity}
\end{equation}
where we have dropped the normalization $C$ and the constant current $j$. 

% local diffusivity
%\begin{figure}[t]
%  \includegraphics[width=0.75\columnwidth]{Diffusivity.eps}
%  \caption{Local diffusivity $D(T)$ (left ordinate) for a random walk in temperature space preformed by a 
%                  replica in a parallel tempering simulation of the chicken villin headpiece subdomain HP-36 
%                  using the ECEPP/2 force field and an implicit solvent.
%                  The diffusivity shows a strong modulation with temperature, note the logarithmic scale of the 
%                  left ordinate. A pronounced minimum in the local diffusivity occurs slightly below the helix-coil 
%                  transition around $T \approx 500$~K where the specific heat $C_V(T)$  (right ordinate) has a 
%                  maximum (dashed line).}
%  \label{Fig:Diffusivity}
%\end{figure}

For the parallel tempering simulation of HP-36 this quantity is plotted in Fig.~\ref{Fig:Diffusivity}. The diffusivity shows a strong modulation along the simulated temperature range $250-1000$~K, note the logarithmic scale of the ordinate. A pronounced minimum occurs around $T\approx 490$~K where the diffusivity is suppressed by two orders of magnitude in comparison to the temperature range below 350~K and above 600~K. This minimum in the diffusivity points to a severe bottleneck for the random walk in temperature space: replicas can move back and forth in temperature rapidly below and above this bottleneck, but experience a dramatic slowdown as they approach and pass through the temperature range around 490~K. This behavior can be explained  through a free energy barrier associated with a structural transition of the protein; the minimum in the local diffusivity is located slightly below the maximum of the specific heat at $T\approx 500$~K which is also plotted in Fig.~\ref{Fig:Diffusivity}. For HP-36 in the ECEPP/2 force-field it has been shown that the position of this peak separates a high-temperature phase with extended unordered configurations from a low-temperature region that is characterized by high helical content of the molecule \cite{Li:03}. 
Below this transition a shoulder in the measured local diffusivity points to a second bottleneck in the simulation for an extended range of temperatures $350~$K$~\leq T \leq 490~$K, possibly caused by competing low-energy configurations with high helical content. While the specific heat for this temperature range is slightly larger than in the high-temperature region above 600~K, there is no characteristic feature similar to the progression of the local diffusivity. 
%Below 350~K, its increase suggests that the replica  wander quickly to low temperatures. 
The local diffusivity is thus a more sensitive measure to identify bottlenecks in a parallel tempering simulation and to locate the multiple temperature scales dominating the folding process of a protein for a given force field.

% Temperature set optimization ---------------------------------------------------------------------

% temperature sets
%\begin{figure}[t]
%  \includegraphics[width=0.75\columnwidth]{TemperatureSets.eps}
%  \caption{(color online) 
%                 Optimized temperature sets for a parallel tempering of HP-36 obtained by the feedback 
%                 algorithm for two different initial temperature sets. 
%                 Independent of the initial temperature set the optimized temperature sets converge to
%                 a temperature set that concentrates temperatures in the vicinity of the helix-coil transition
%                 temperature around $T\approx 500$~K (dashed line). }
%  \label{Fig:T-Sets}
%\end{figure}

In order to speed up equilibration we want to maximize the rate of round-trips which each replica performs between the two extremal temperatures, or equivalently the diffusive current $j$, by varying the temperature set $\{ T_i \}$ and thus the probability distribution $\eta(T)$, as discussed in Refs.~\cite{OptimalEnsemble,OptimalTempering}. Rearranging and integrating Eq.~(\ref{Eq:Current}) this goal is achieved by minimizing the integral
\begin{equation}
\frac{1}{j} = \int \left[ \frac{1}{D(T)\eta(T)} + \lambda\eta(T) \right] dT \;,
\label{Eq:integral}
\end{equation}
where we have added a Lagrange multiplier $\lambda$ which ensures that $\eta(T)$ remains a normalized probability distribution. Varying the probability distribution $\eta(T)$ the integrand in 
Eq.~(\ref{Eq:integral}) is minimized for 
\begin{equation}
  \eta^{\rm(opt)}(T) = \frac{C'}{\Delta T'} = C' \sqrt{ \frac{1}{\Delta T} \, \frac{df}{dT} } \sim \frac{1}{\sqrt{D(T)}} \;,
\end{equation}
where the normalization $C'$ is again chosen according to the normalization condition in Eq.~(\ref{Eq:Normalization}).  
For the optimal temperature set the temperature points are thus rearranged in such a way that the probability distribution $\eta^{\rm(opt)} (T)$ becomes inversely proportional to the square root of the local diffusivity.
Measuring the local diffusivity $D(T)$ for an initial temperature set, we can determine the optimized probability distribution $\eta^{\rm (opt)}(T)$ approximated as a step-function in the original temperature set. The {\em optimized} temperature set $\{ T'_i\}$ is then found by choosing the $n$-th temperature 
point $T'_n$ such that
\begin{equation}
  \int_{T'_1}^{T'_n} \eta^{\rm (opt)}(T) dT = n/N \;,
\end{equation}
where $1<n<N$ and the two extremal temperatures $T'_1=T_1$ and $T'_N=T_N$ remain fixed.
This feedback of the local diffusivity is then iterated for increasingly long simulation runs 
-- in our simulations we double the number of swaps for subsequent feedback steps --
until convergence of the optimized temperature set is found.

% acceptance probabilities
%\begin{figure}[t]
%  \includegraphics[width=0.75\columnwidth]{AcceptanceProbability.eps}
%  \caption{Acceptance probabilities (open squares) of replica swaps in a parallel tempering simulation 
%  	         of HP-36 using the optimized temperature set illustrated in Fig.~\ref{Fig:T-Sets}.
%	         The dependence of the acceptance probabilities on the temperature closely reflects the shape 
%	         of the measured local diffusivity (filled circles). 
%                  In the vicinity of the helix-coil transition temperature $T\approx 500$~K where the local
%                  diffusivity is strongly suppressed the acceptance probabilities are highest
%                  due to the contraction of temperature points in the optimized temperature set.
%                  The dotted lines indicates the minimum in the local diffusivity.}
%  \label{Fig:Pacc}
%\end{figure}

In our simulations we start with the arbitrary initial temperature set of Table \ref{Table:TSets} 
that similar to a geometric progression concentrates temperature points at low temperatures. 
Three feedback steps were performed, one after 100,000 MC sweeps, a second after further 
200,000 sweeps, and a third one after additional 400,000 sweeps. The iterated temperature sets are plotted in Fig.~\ref{Fig:T-Sets} and also 
listed in Table \ref{Table:TSets}. 
The feedback algorithm shifts computational resources towards the temperature of the 
helix-coil transition and temperature points in the optimized temperature sets concentrate 
around $T\approx 500$~K where the measured local diffusivity is suppressed, see 
Fig.~\ref{Fig:Diffusivity}.
In the derivation of the feedback procedure we have assumed that the local diffusivity is to
leading order independent from the temperature set. A posteriori we can verify this 
assumption by demonstrating that the optimized temperature set is independent of the initial
temperature set. To this end, we perform a second series of feedback optimization steps 
starting from the temperature set of Ref.~\onlinecite{Li:03}. As illustrated in the lower half of  
Fig.~\ref{Fig:T-Sets} we indeed find that a very similar distribution is approached.

%\begin{table*}[t]
i%  \caption{Temperature sets used in the parallel tempering simulation of HP-36.
%                  Applying the feedback algorithm temperature points in the optimized temperature
%                  sets (iterations 2,3 and 4) concentrate around the helix-coil transition around 500~K.
%                  The temperature sets are also illustrated in Fig.~\ref{Fig:T-Sets}.}
%  \begin{tabular}{c | llllllllllllllllllll}
%  \hline
%iteration & \multicolumn{20}{c}{temperature set [$K$]} \\
%  \hline \hline
%1 & 
%250 & 275 & 300 & 325 & 350 & 375 & 400 & 425 & 450 &
%500 & 550 & 600 & 650 & 700 & 750 & 800 & 850 & 900 & 950 & 1000 \\
% \hline
% 2 &
%250 & 295 & 326 & 349 & 371 & 395 & 424 & 446 & 464 &
%482 & 499 & 514 & 528 & 543 & 559 & 577 & 595 & 628 & 693 & 1000 \\
%  \hline
%3 & 
%250 & 326 & 359 & 385 & 411 & 434 & 452 & 467 & 480 &
%491 & 501 & 510 & 519 & 527 & 536 & 546 & 560 & 578 & 619 & 1000 \\
%   \hline
%4 & 
%250 & 314 & 358 & 381 & 402 & 423 & 444 & 461 & 474 &
%484 & 494 & 502 & 511 & 519 & 529 & 540 & 554 & 576 & 670 & 1000 \\
%   \hline
%  \end{tabular}
%  \label{Table:TSets}
%\end{table*}

For the optimized temperature set  the acceptance probabilities of replica swaps show a strong {\em temperature dependence} as illustrated in Fig.~\ref{Fig:Pacc}. This is a consequence of 
the concentration of temperature points around $T\approx 500$~K in the optimized temperature set for HP-36. There the acceptance probabilities are found to be relatively high (around 80\%) while in the temperature regions below 350~K and above 600~K where temperature points have been thinned out the acceptance probabilities drop below some 40\%.
The fact that for our optimized temperature set the acceptance probabilities vary with temperature contradicts various alternative approaches in the literature \cite{Kofke:04,Predescu:04,Kone:05,Rathore:05,Predescu:05} that aim at maximizing equilibration by choosing a temperature set where the acceptance probability of attempted swaps is independent of temperature. 

% HP-36 results ---------------------------------------------------------

% radius of gyration
%\begin{figure}[t]
%  \includegraphics[width=0.75\columnwidth]{Rgy.eps}
%  \caption{(color online) Radius of gyration of the lowest-energy configuration in a parallel tempering 
%                  simulation of HP-36 versus the number of Monte Carlo sweeps.
%                  The dashed lines indicate when the temperature set used in the simulation was redefined 
%                  as illustrated in the upper half of Fig.~\ref{Fig:T-Sets}.
%                  The insets show histograms of the radius of gyration for the three simulation parts.}
%  \label{Fig:RGY}
%\end{figure}

\section{Simulation results --}
The feedback-iterations systematically optimize the temperature set which maximize the 
efficiency of parallel tempering simulations. We now turn to the results obtained for our 
simulations of HP-36 and discuss the effects of the temperature reweighting.
Though the parallel tempering simulations allow to evaluate thermodynamic quantities over a range of temperatures, here we focus on the properties of the configurations at the lowest temperatures. In Fig.~\ref{Fig:RGY} the radius of gyration $R_{gy}$ which measures the compactness of a protein configuration is plotted for the lowest-energy configuration versus the number of Monte Carlo sweeps. For the initial iteration 
%where we used the temperature set of Ref.~\cite{Li:03} 
 the radius of gyration varies in a broad range of $10-14$~\AA. A histogram of $R_{gy}$ is plotted on top of the time series in Fig.~\ref{Fig:RGY} showing that two sets of configurations are found, one set with "compact" configurations characterized by a radius of gyration in the range $10-11$~\AA\  and "extended" configurations with a radius of gyration in the range $12-14$~\AA. Averaging over some 100,000 MC sweeps in the first  iteration we find that about 25\% of the configurations are "compact", and a remaining 75\% of "extended" configurations. Previous simulations \cite{Li:03} 
with a total of 150,000 MC sweeps  also reported the occurrence of these two sets of configurations. 
Similar to our case the "extended" configurations dominated, and only a small fraction of 20\% of the configurations were "compact". 
%Time series of the potential energy suggested that those simulations had reached equilibrium and therefore the measurement of the distribution correct. 
% Figure: Time series of energy. Inlay distribution of energy over radius of gyration
In the present study we continued the simulation after the first feedback step  with an optimized temperature set for an additional 200,000 MC sweeps. The time series in Fig.~\ref{Fig:RGY} shows that as a consequence, the fraction of "extended" configurations in the lowest-energy configurations is significantly reduced and some 90\% of the sampled lowest-energy configuration have a radius of gyration smaller than 11~\AA. This ratio increases further to 99\% for the final iteration 
with 400,000 MC sweeps after the second feedback step. 
While in the previous study equilibration at low temperatures was determined by analyzing the time series for thermodynamic observables such as the potential energy and convergence was found after some 100,000 MC sweeps, the discrepancy to the results presented here cast serious doubt whether an overall simulation time of some 150,000 MC sweeps and a sub-optimal temperature set were sufficient to reach full equilibration.
The long relaxation times in our example indicate that even with a sophisticated technique
like parallel tempering the simulation times have to be considerably longer than commonly assumed.
In order to assure equillibrisation at lowest temperature the number of round trip times should be
at least $n_{rt} \approx 10$.
%Previous simulations \cite{Li:03} 
%with the same temperature set used in iteration 1, but a smaller number of 
%with 
%total MC sweeps (150,000) also reported the occurrence of these two sets of configurations. In the time series obtained in those simulations the "extended" configurations dominated, and only a small fraction of 20\% of the configurations were "compact". While thermodynamic observables such as the potential energy suggested that those simulations were equilibrating at low temperatures, the discrepancy to the results presented here cast serious doubt whether an overall simulation time of some 150,000 MC sweeps and a sub-optimal temperature set were sufficient to reach full equilibration.

% lowest-energy configuration
%\begin{figure}[t]
%  \includegraphics[width=0.7 \columnwidth]{hp36-emin.eps}
%  \caption{(color online) Lowest energy structure of HP-36 as obtained by an all-atom 
%                 Monte Carlo simulation using the ECEPP/2 force field and an implicit solvent.
%                 The root-mean square deviation of this structure to the PDB structure shown 
%                 in Fig.~\ref{Fig:HP36} is $r_{\text{RMSD}}=3.7$~\AA.}
%  \label{Fig:LowestEnergy-Configuration}
%\end{figure}

To probe whether our simulations allow a structural prediction of the true ground state configuration we track the configuration with the overall lowest energy in the simulation and compare it to the Protein Data Bank structure of HP-36 (PDB code 1vii).
The lowest-energy configuration obtained in our simulation is illustrated in Fig.~\ref{Fig:LowestEnergy-Configuration}. Despite the fact that in this structure the two N-terminal helices merged to one long
helix (compromising residues 5 to 21) that tightly packs to the C-terminal helix, its  RMSD to the PDB structure is only  $r_{\text{RMSD}}=3.7$~\AA. This value is substantially lower than in the structures with an RMSD of $r_{\text{RMSD}} \approx 5.8 - 6.0$~\AA\  previously obtained by molecular dynamics simlations \cite{Duan:98}, Monte Carlo simulations \cite{Li:03,Hansmann:04} and optimization techniques \cite{Hansmann:02}.  A structure with comparable RMSD of  $r_{\text{RMSD}}\approx3.5$~\AA\  has been obtained by large-scale molecular dynamics simulations \cite{Zagrovic:02}. However, in those simulations the best-matching structure was found by comparing {\em all} sampled configurations along multiple trajectories to the PDB structure, while in our simulations the optimal structure is singled out as the one with the lowest energy. In addition, our simulations consumed only about 1\% of the computing time resources (about 1,000 cpu years) used in Ref.~\onlinecite{Zagrovic:02}.

% Summary / Conclusions ----------------------------------------------------

\section{Conclusions --}
In conclusion, we have applied a powerful feedback algorithm for the numerical simulation
of proteins that allows to allocate computational resources in a parallel tempering simulation
so that equilibration at all temperatures is considerably improved.
By tracking the diffusion of replicas in temperature space we have identified the bottlenecks of 
a simulation, typically in the vicinity of the folding transition. Feeding back this information we 
obtain an optimal temperature set that concentrates temperature points at these bottlenecks. 
Our algorithm differs from previous approaches that aim at maximizing equilibration by 
considering the local acceptance probabilities of replica exchange moves. In contrast we find
that for the optimal temperature set acceptance probabilities for such swap moves show a strong
temperature dependence. 
Applying the optimized parallel tempering technique to the simulation of the 36-residue protein
HP-36 we find a dominant low-energy configuration with less than 4 \AA\  root-mean square 
distance from the native structure within a fraction of the computing time consumed by 
high-performance molecular dynamics simulations. 

We note, however, that the energy difference between our compact, lowest-energy configuration 
and the extended structure with lowest energy -- which differs from the PDB structure by
an RMSD of 8.0 \AA -- is only $\approx 10$ kcal/mol (for the minimized configurations).
On the other hand, the energy of our lowest-energy configuration is 100 kcal/mol {\it lower} than 
that of the (minimized) PDB structure from which (despite the small RMSD) it still differs 
considerably. 
Hence, while our results appear to be closer to the experimental results than previous simulations
they still demonstrate the limitations on protein simulations that are inherent 
in present energy functions. The extremely long relaxation times indicate the existence of spurious minima that should be absent in the folding funnel of fast folding proteins such as the villin headpiece. 
Unveiling these limitations in the energy functions and their underlying causes requires optimized simulation techniques such as the one applied in the present paper.  

% Acknowledgments ---------------------------------------------------------

{\em Acknowledgments --}
We thank A. Laio and B. Zagrovic  for stimulating discussions, as well as D. A. Huse and 
H. G. Katzgraber for fruitful discussions on the technical aspects of this manuscript. 
S.T. acknowledges support by the Swiss National Science Foundation, U.H. by a
research grant (CHE-9981874) of the National Science Foundation (USA).

\newpage
% References

\newpage
\setcounter{table}{0}
\begin{table*}[t]
  \caption{Temperature sets used in the parallel tempering simulation of HP-36.
                  Applying the feedback algorithm temperature points in the optimized temperature
                  sets (iterations 2,3 and 4) concentrate around the helix-coil transition around 500~K.
                  The temperature sets are also illustrated in Fig.~\ref{Fig:T-Sets}.}
  \begin{tabular}{c | llllllllllllllllllll}
  \hline
iteration & \multicolumn{20}{c}{temperature set [$K$]} \\
  \hline \hline
1 & 
250 & 275 & 300 & 325 & 350 & 375 & 400 & 425 & 450 &
500 & 550 & 600 & 650 & 700 & 750 & 800 & 850 & 900 & 950 & 1000 \\
 \hline
 2 &
250 & 295 & 326 & 349 & 371 & 395 & 424 & 446 & 464 &
482 & 499 & 514 & 528 & 543 & 559 & 577 & 595 & 628 & 693 & 1000 \\
  \hline
3 & 
250 & 326 & 359 & 385 & 411 & 434 & 452 & 467 & 480 &
491 & 501 & 510 & 519 & 527 & 536 & 546 & 560 & 578 & 619 & 1000 \\
   \hline
4 & 
250 & 314 & 358 & 381 & 402 & 423 & 444 & 461 & 474 &
484 & 494 & 502 & 511 & 519 & 529 & 540 & 554 & 576 & 670 & 1000 \\
   \hline
  \end{tabular}
  \label{Table:TSets}
\end{table*}
%%% figure legends

\clearpage

\begin{figure}[!h]
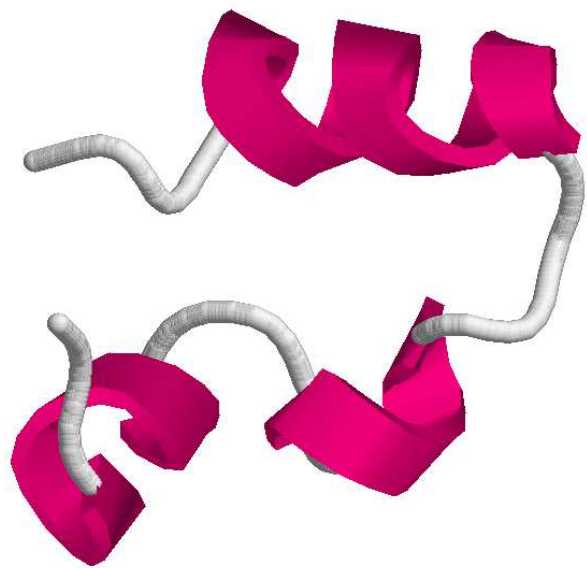

  \caption{(color online) NMR-derived structure of the 36-residue peptide HP-36 
                   as deposited in the Protein Data Bank (1vii).}
\end{figure}

\begin{figure}[!h]
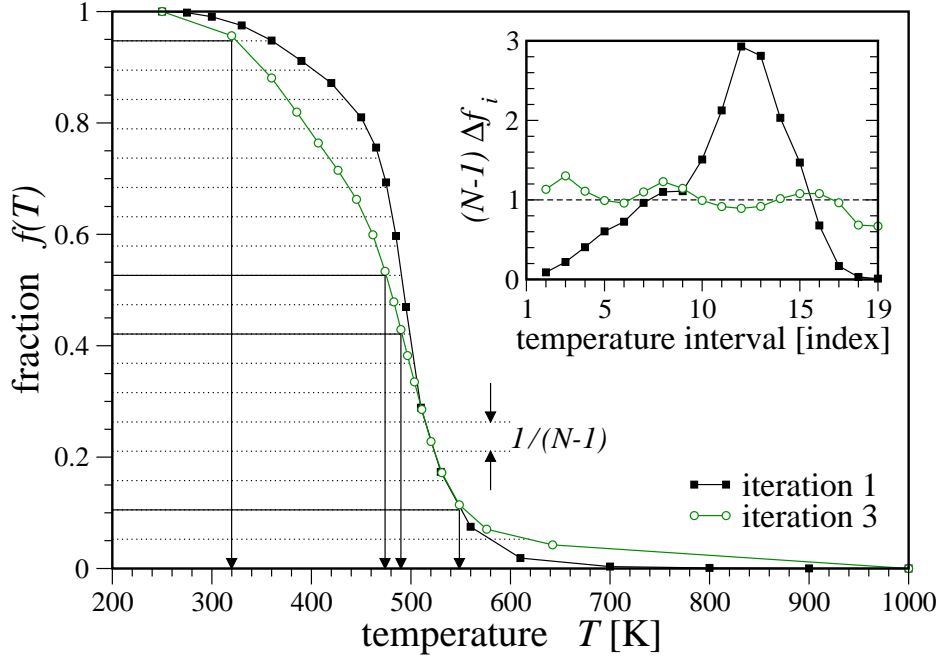

  \caption{Fraction of replicas diffusing from the lowest, $T_1=250~$K, to the highest temperature
                  $T_N=1000~$K in a parallel tempering simulation of HP-36.
                  For the optimized temperature set (iteration 3), the temperature points are distributed in
                  such a way that the fraction shows a nearly constant decay 
                  $\Delta f_i = f(T_i) - f(T_{i+1})=1/(N-1)$ between adjacent temperature points,
                  i.e. $\Delta f_i (N-1) \approx const$.}
\end{figure}

\begin{figure}[!h]
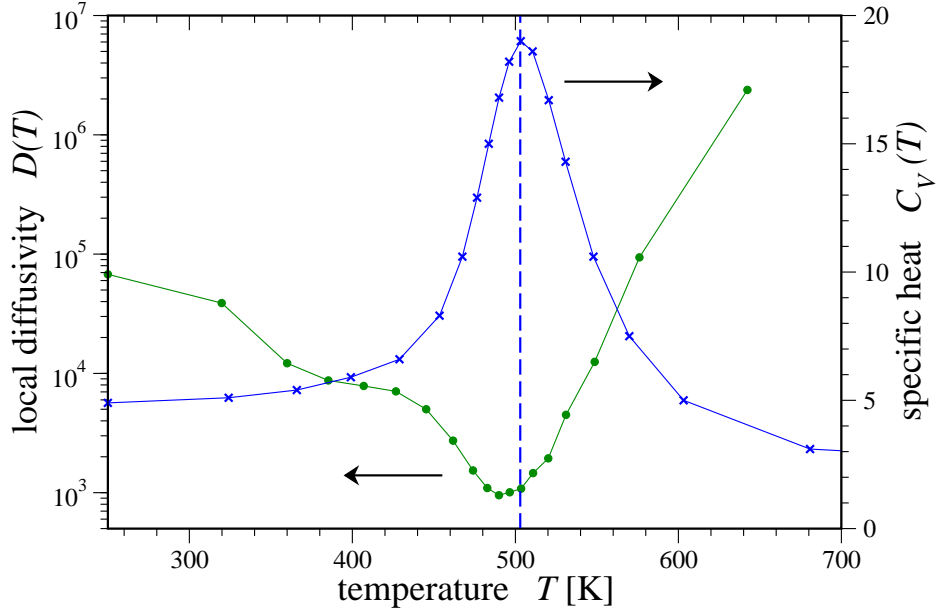

  \caption{Local diffusivity $D(T)$ (left ordinate) for a random walk in temperature space preformed by a 
                  replica in a parallel tempering simulation of the chicken villin headpiece subdomain HP-36 
                  using the ECEPP/2 force field and an implicit solvent.
                  The diffusivity shows a strong modulation with temperature, note the logarithmic scale of the 
                  left ordinate. A pronounced minimum in the local diffusivity occurs slightly below the helix-coil 
                  transition around $T \approx 500$~K where the specific heat $C_V(T)$  (right ordinate) has a 
                  maximum (dashed line).}
\end{figure}

\begin{figure}[!h]
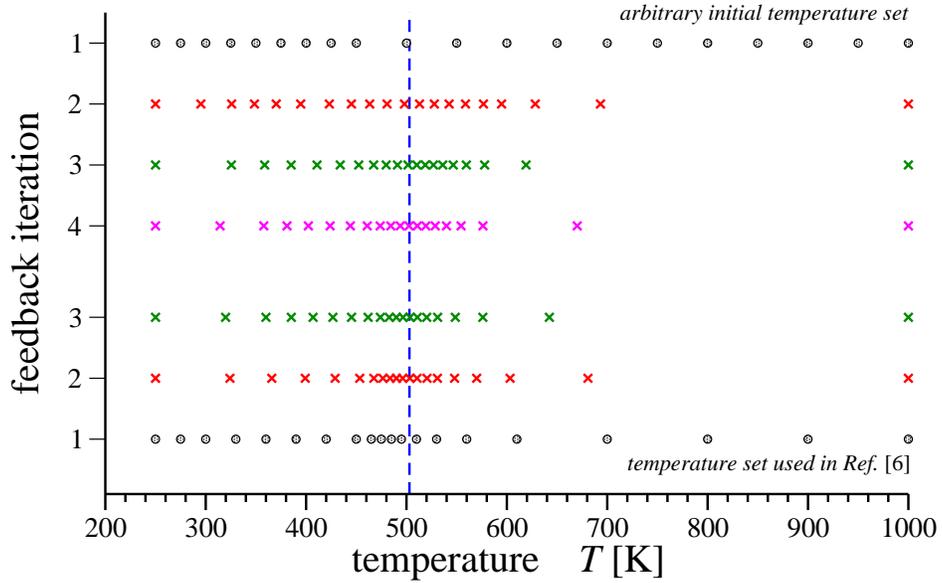

  \caption{(color online) 
                 Optimized temperature sets for a parallel tempering of HP-36 obtained by the feedback 
                 algorithm for two different initial temperature sets. 
                 Independent of the initial temperature set the optimized temperature sets converge to
                 a temperature set that concentrates temperatures in the vicinity of the helix-coil transition
                 temperature around $T\approx 500$~K (dashed line). }
\end{figure}

\begin{figure}[!h]
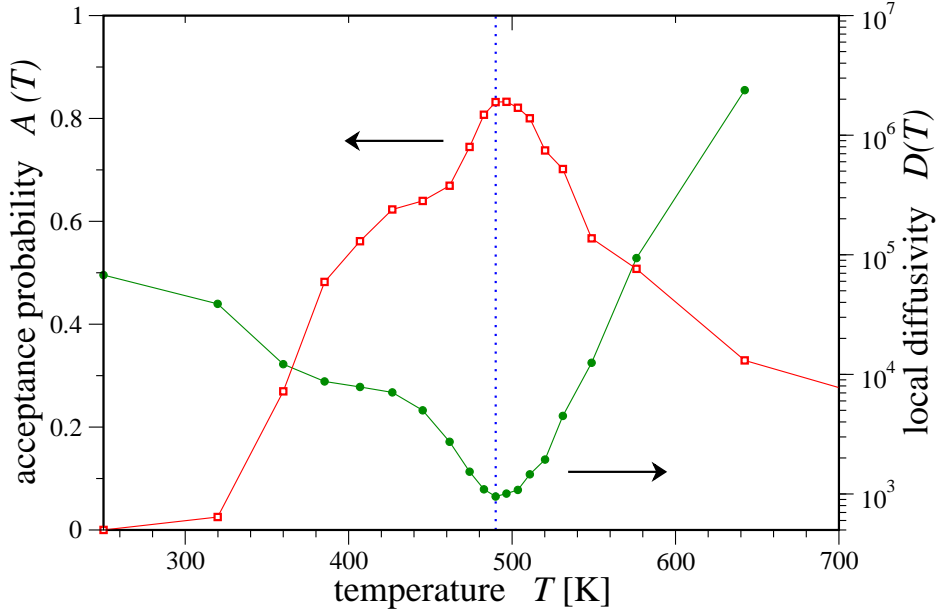

  \caption{Acceptance probabilities (open squares) of replica swaps in a parallel tempering simulation 
  	         of HP-36 using the optimized temperature set illustrated in Fig.~\ref{Fig:T-Sets}.
	         The dependence of the acceptance probabilities on the temperature closely reflects the shape 
	         of the measured local diffusivity (filled circles). 
                  In the vicinity of the helix-coil transition temperature $T\approx 500$~K where the local
                  diffusivity is strongly suppressed the acceptance probabilities are highest
                  due to the contraction of temperature points in the optimized temperature set.
                  The dotted lines indicates the minimum in the local diffusivity.}
\end{figure}

\begin{figure}[!h]
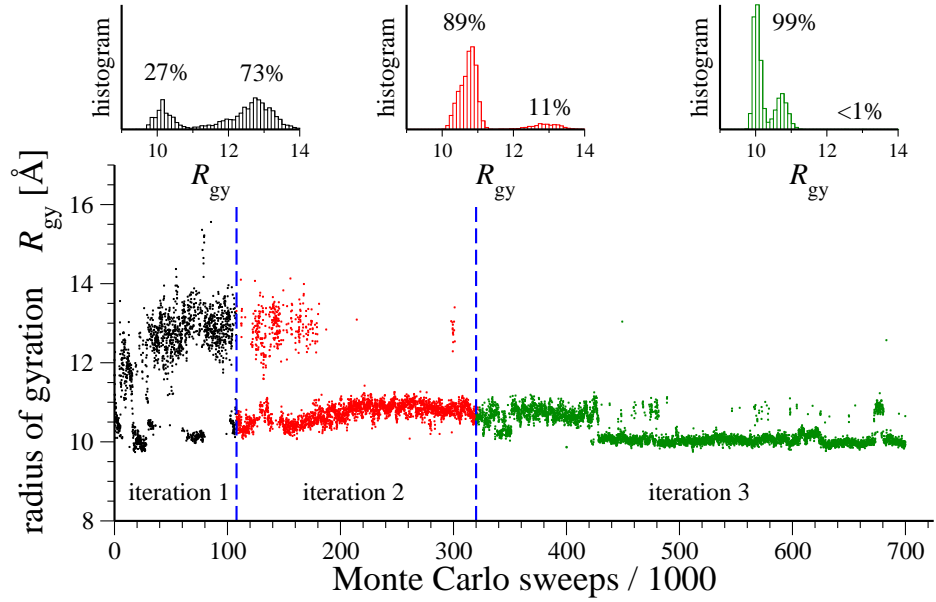

  \caption{(color online) Radius of gyration of the lowest-energy configuration in a parallel tempering 
                  simulation of HP-36 versus the number of Monte Carlo sweeps.
                  The dashed lines indicate when the temperature set used in the simulation was redefined 
                  as illustrated in the upper half of Fig.~\ref{Fig:T-Sets}.
                  The insets show histograms of the radius of gyration for the three simulation parts.}
\end{figure}

\begin{figure}[!h]
  \caption{(color online) Lowest energy structure of HP-36 as obtained by an all-atom 
                 Monte Carlo simulation using the ECEPP/2 force field and an implicit solvent.
                 The root-mean square deviation of this structure to the PDB structure shown 
                 in Fig.~\ref{Fig:HP36} is $r_{\text{RMSD}}=3.7$~\AA.}
\end{figure}

\clearpage
\setcounter{figure}{0}
% NMR configuration
 \begin{figure}[b]
   \includegraphics[width=0.5\columnwidth]{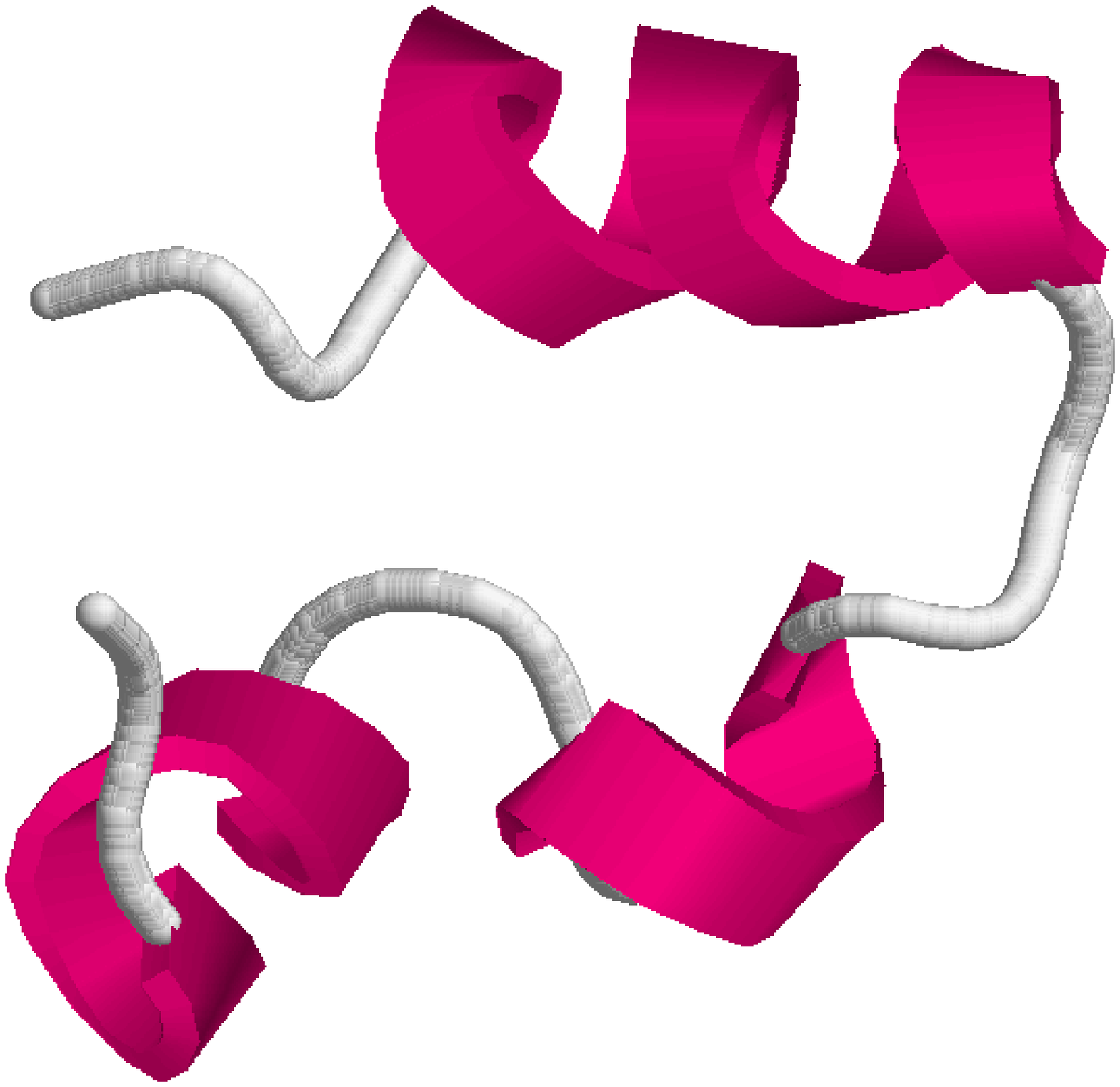}
   \caption{(color online) NMR-derived structure of the 36-residue peptide HP-36 
                    as deposited in the Protein Data Bank (1vii).}
   \label{Fig:HP36}
 \end{figure}
\clearpage

% fraction & feedback
\begin{figure}[t]
  \includegraphics[width=0.75\columnwidth]{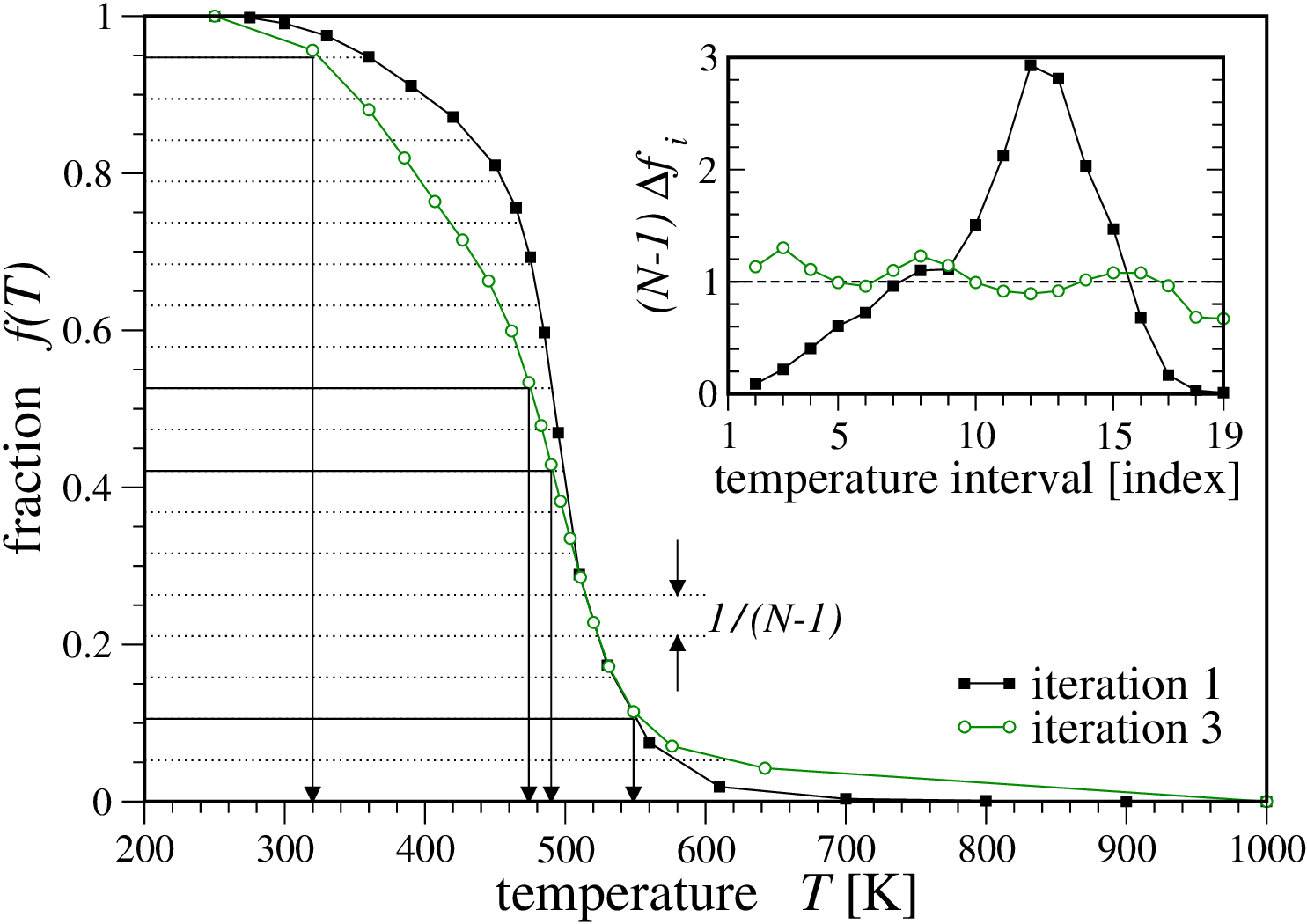}
  \caption{Fraction of replicas diffusing from the lowest, $T_1=250~$K, to the highest temperature
                  $T_N=1000~$K in a parallel tempering simulation of HP-36.
                  For the optimized temperature set (iteration 3), the temperature points are distributed in
                  such a way that the fraction shows a nearly constant decay 
                  $\Delta f_i = f(T_i) - f(T_{i+1})=1/(N-1)$ between adjacent temperature points,
                  i.e. $\Delta f_i (N-1) \approx const$.}
  \label{Fig:Fraction}
\end{figure}

\clearpage

% local diffusivity
\begin{figure}[t]
  \includegraphics[width=0.75\columnwidth]{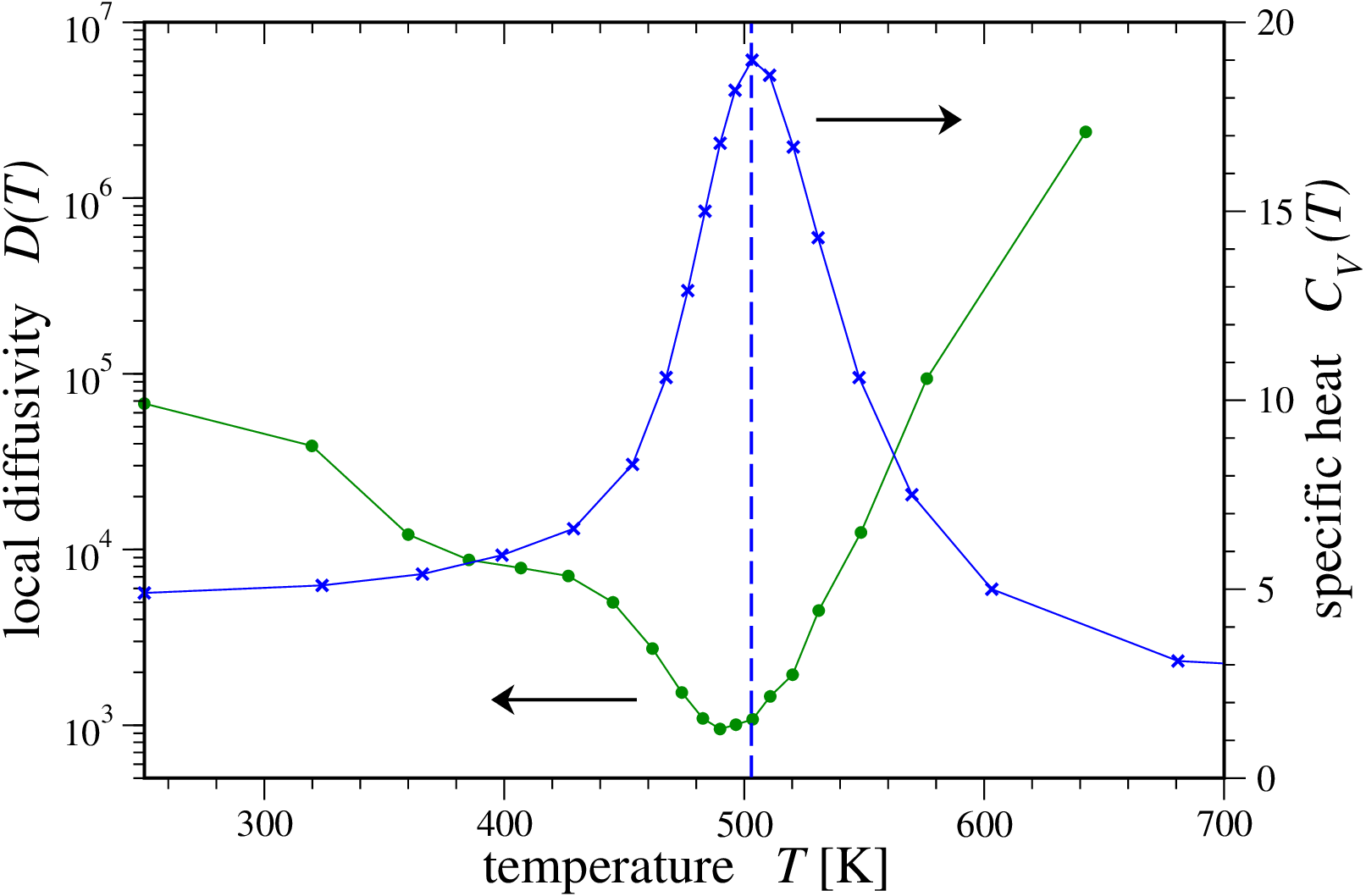}
  \caption{Local diffusivity $D(T)$ (left ordinate) for a random walk in temperature space preformed by a 
                  replica in a parallel tempering simulation of the chicken villin headpiece subdomain HP-36 
                  using the ECEPP/2 force field and an implicit solvent.
                  The diffusivity shows a strong modulation with temperature, note the logarithmic scale of the 
                  left ordinate. A pronounced minimum in the local diffusivity occurs slightly below the helix-coil 
                  transition around $T \approx 500$~K where the specific heat $C_V(T)$  (right ordinate) has a 
                  maximum (dashed line).}
  \label{Fig:Diffusivity}
\end{figure}
\clearpage

% temperature sets
\begin{figure}[t]
  \includegraphics[width=0.75\columnwidth]{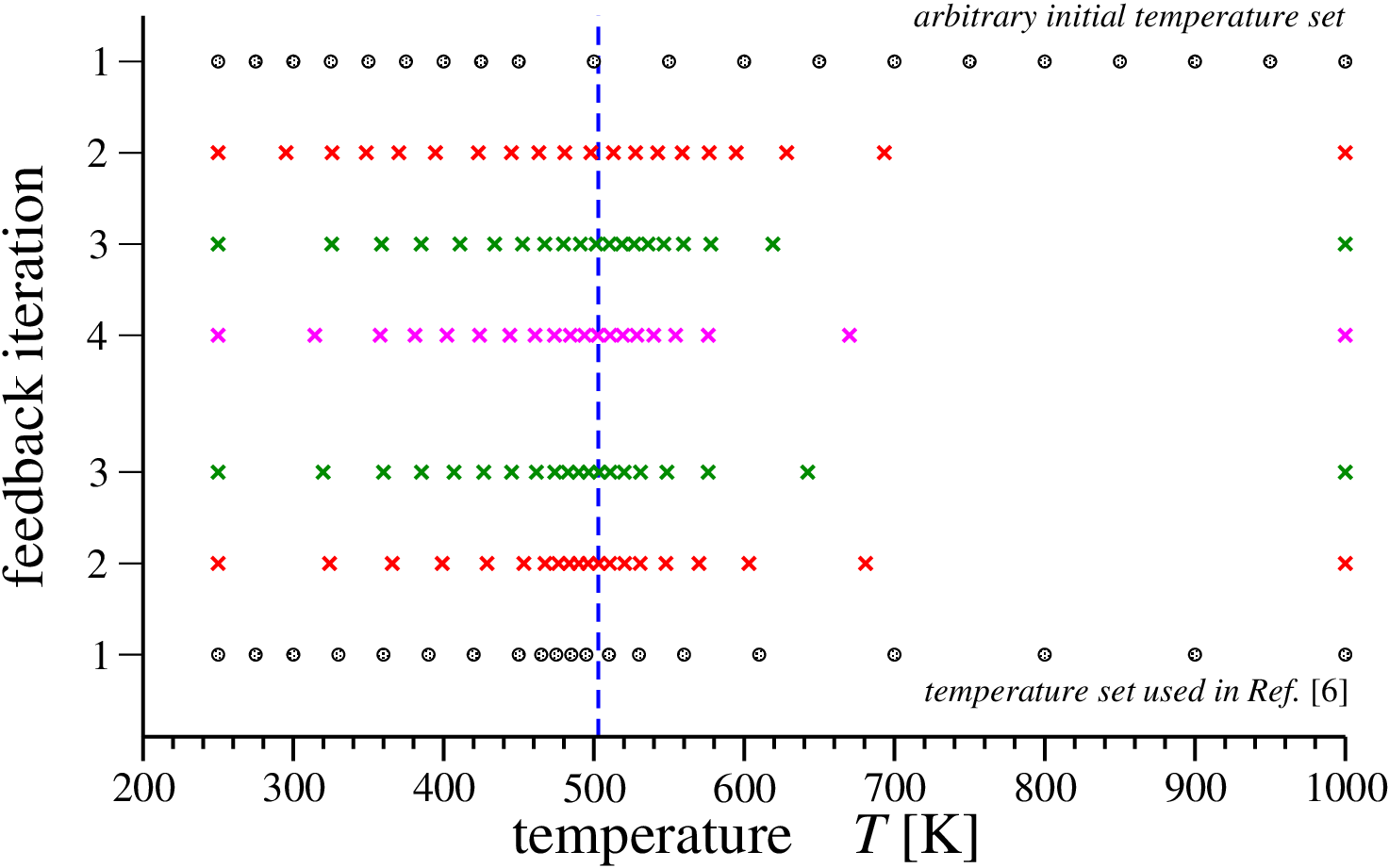}
  \caption{(color online) 
                 Optimized temperature sets for a parallel tempering of HP-36 obtained by the feedback 
                 algorithm for two different initial temperature sets. 
                 Independent of the initial temperature set the optimized temperature sets converge to
                 a temperature set that concentrates temperatures in the vicinity of the helix-coil transition
                 temperature around $T\approx 500$~K (dashed line). }
  \label{Fig:T-Sets}
\end{figure}
\clearpage

% acceptance probabilities
\begin{figure}[t]
  \includegraphics[width=0.75\columnwidth]{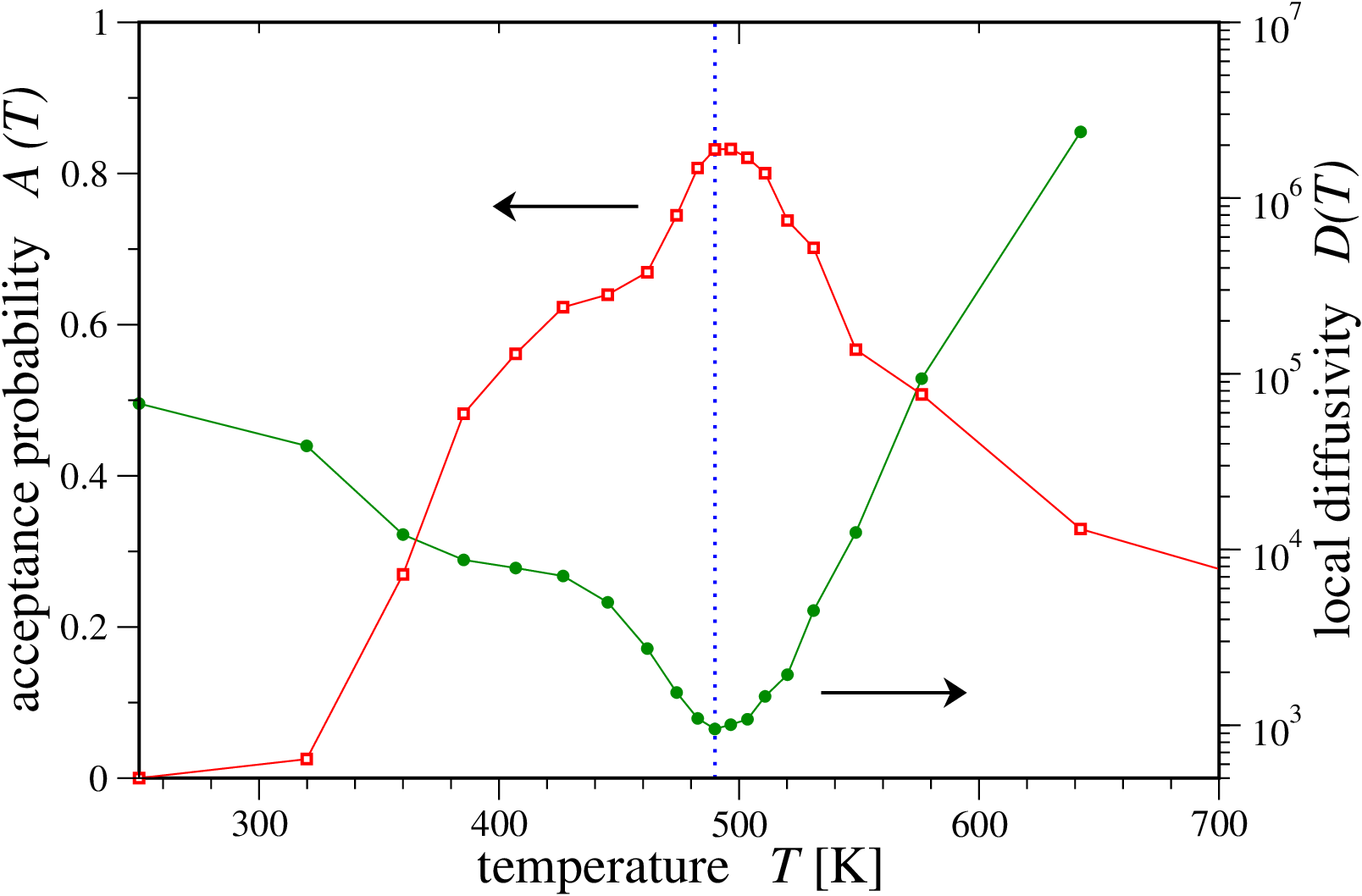}
  \caption{Acceptance probabilities (open squares) of replica swaps in a parallel tempering simulation 
  	         of HP-36 using the optimized temperature set illustrated in Fig.~\ref{Fig:T-Sets}.
	         The dependence of the acceptance probabilities on the temperature closely reflects the shape 
	         of the measured local diffusivity (filled circles). 
                  In the vicinity of the helix-coil transition temperature $T\approx 500$~K where the local
                  diffusivity is strongly suppressed the acceptance probabilities are highest
                  due to the contraction of temperature points in the optimized temperature set.
                  The dotted lines indicates the minimum in the local diffusivity.}
  \label{Fig:Pacc}
\end{figure}
\clearpage

% radius of gyration
\begin{figure}[t]
  \includegraphics[width=0.75\columnwidth]{Rgy.eps}
  \caption{(color online) Radius of gyration of the lowest-energy configuration in a parallel tempering 
                  simulation of HP-36 versus the number of Monte Carlo sweeps.
                  The dashed lines indicate when the temperature set used in the simulation was redefined 
                  as illustrated in the upper half of Fig.~\ref{Fig:T-Sets}.
                  The insets show histograms of the radius of gyration for the three simulation parts.}
  \label{Fig:RGY}
\end{figure}

\clearpage

% lowest-energy configuration
\begin{figure}[t]
  \includegraphics[width=0.7 \columnwidth]{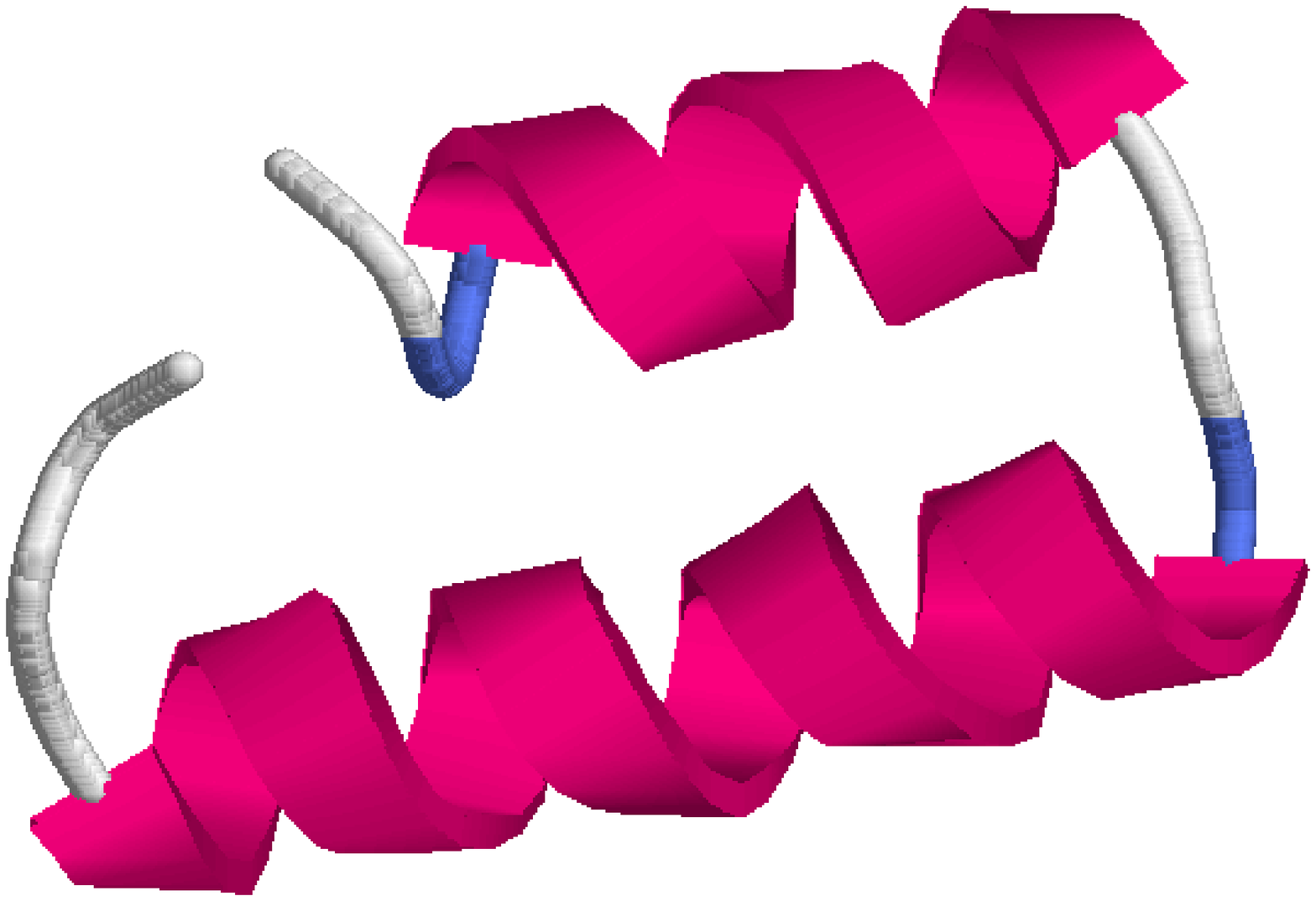}
  \caption{(color online) Lowest energy structure of HP-36 as obtained by an all-atom 
                 Monte Carlo simulation using the ECEPP/2 force field and an implicit solvent.
                 The root-mean square deviation of this structure to the PDB structure shown 
                 in Fig.~\ref{Fig:HP36} is $r_{\text{RMSD}}=3.7$~\AA.}
  \label{Fig:LowestEnergy-Configuration}
\end{figure}

\end{document}